\documentclass[prc,twocolumn,preprintnumbers,superscriptaddress,showpacs,nofootinbib]{revtex4-1}
\usepackage{graphicx,times}
\usepackage{amsmath,amsfonts,amssymb}

\begin{document}

\title{$K\Lambda(1405)$ configuration of the $K\bar{K}N$ system}

\author{A. Mart\'inez Torres}
\author{D. Jido}
\affiliation{
Yukawa Institute for Theoretical Physics, Kyoto University,
Kyoto 606-8502, Japan}
\preprint{YITP-10-68}

\date{\today}

\begin{abstract}
We study the $K\Lambda(1405)$ configuration of the $K\bar{K}N$ system by considering $K\pi\Sigma$ as a coupled channel. We solve the Faddeev equations for these systems and find confirmation of the existence of a new $N^{*}$ resonance around 1920 MeV with $J^{\pi}=1/2^{+}$ predicted in a single-channel potential model and also found in a Faddeev calculation as an $a_{0}(980)N$  state, with the $a_{0}(980)$ generated in the $K\bar{K}$, $\pi\eta$ interaction.
\end{abstract}

\maketitle


The study of three-body systems is one of the important issues 
of contemporary nuclear physics and has attracted continuous attention.
Chronicled examples are baryonic three-body bound systems, such as 
tritium, $^{3}$He ($NNN$) and hyper triton ($pn\Lambda$). 
Recently, interest in three-body systems has developed,
and resonance systems including mesons as the constituents are 
considered based on current knowledge of hadronic interaction.

Evidences for several existing and new states which can be interpreted as three-body resonances are being reported from theoretical and experimental studies.
For example,  it has been claimed that the $Y(4660)$ resonance found in 
$ e^{+}e^{-}\to \gamma_{ISR}  \pi^{+ }\pi^{-}\psi^{\prime}$ can be interpreted 
as a $f_{0}(980)\psi^{\prime}$ bound state~\cite{guo}. 
For the $\pi\bar{K} N$ system,
the Faddeev equations were solved using unitary chiral dynamics and 
coupled channels and dynamical generation of all the $\Sigma$ and $\Lambda$ 
resonance states with $J^{P}=1/2^{+}$ listed by the Particle Data Group (PDG) 
in the energy region 1500-1800 MeV was found in Ref.~\cite{mko2}. 
The same formalism applied to the $\pi\pi N$ system and its coupled channels revealed 
the dynamical generation of the $N^{*}(1710)$, $N^*(2100)$ and 
$\Delta(1910)$~\cite{mko3}. 
The $X(2175)$ state, reported by different experimental groups~\cite{babar1,bes} 
in the $\phi f_{0}(980)$ invariant mass, has been explained as a $\phi K\bar{K}$ resonance with $K\bar{K}$ forming the $f_{0}(980)$ resonance~\cite{mko5,oller3,eef}.  

Further, kaonic nuclear few-body systems are of special interest 
in relation with strangeness nuclear physics. 
Possible existence of $\bar KNN$ bound states was pointed out in 60's~\cite{nogami}
by considering the $\Lambda(1405)$ hyperon resonance as a quasibound state
of $\bar KN$ as suggested in Ref.~\cite{Dalitz}\footnote{A recent investigation
 using a coupled channel approach based on chiral dynamics also confirmed 
 that the $\Lambda(1405)$ resonance can be described substantially 
 by a meson-baryon molecular state~\cite{Hyodo:2008xr}.}. 
Recently thorough theoretical 
investigations of the $\bar{K}NN$ system in various 
approaches~\cite{yama,nina,ikeda,dote,Wycech:2008wf} indicate a quasibound state 
with a large width. Baryonic systems with two kaons were also investigated 
in Refs.~\cite{enyo,jido} with a single channel variational method. 
For the $K\bar KN$ system~\cite{jido},
a quasibound state of these hadrons was found around 1910 MeV for an $N^{*}$
with $I=1/2$ and $J^{P}$=$1/2^{+}$ using the effective $\bar KN$ 
potential derived in Ref.~\cite{hyodo07} and the $\bar KK$ interactions 
reproducing $f_{0}(980)$ and $a_{0}(980)$ as $\bar KK$ quasibound 
states with 980 MeV mass and 60 MeV width.
In case of the $N^{*}$ state found in Ref. \cite{jido} the $\bar{K}N$ pair forms the $\Lambda(1405)$ 
and, simultaneously, the $K\bar{K}$ pair is resonating as the $a_{0}(980)$. 
The same state was also found independently in a study of 
the $NK\bar{K}$, $N\pi\pi$ and $N\pi\eta$ coupled channels 
based on solution of the Faddeev equations \cite{mko1}. 
There it was concluded that a state indeed appears around 1920 MeV 
when the hadrons rearrange themselves to form a $a_{0}(980)N$ system 
and that the contribution of the $N\pi\pi$ and $N\pi\eta$ channels was negligible 
in the dynamical generation of this resonance. 
A $N^{*}$ state with these properties is not listed  by the PDG, however, 
there have been some proposals that this state can be seen in the data for 
the $\gamma p\to K^{+}\Lambda$ reaction \cite{mart,mko6}, although the situation 
is still controversial~\cite{glander,bradford,sumihama}. 

The single channel variational approach~\cite{jido} found that 
this new $N^{*}$ state substantially contains $\Lambda(1405)$ in the $\bar KN$ subsystem.
In the Faddeev analysis~\cite{mko1}, 
although two-body coupled channels were fully considered, explicit three-body 
channels of $K\pi\Sigma$ and $K\pi\Lambda$ were not taken into account. 
The $K\pi\Sigma$ channel could produce some changes on the characteristics of the $N^{*}$ state found in Ref. \cite {jido, mko1},
since the $\Lambda(1405)$ is dynamically generated in coupled
channels such as $\bar KN$ and $\pi \Sigma$~\cite{kaiser,eo1,oller4,jido2,Hyodo03},
especially the lower pole\footnote{It was pointed out in Refs~\cite{fink,oller4,jido2,jido3} 
 that there exist two poles around 
 the $\Lambda(1405)$ energy region, having them
 different coupling nature to meson-baryon states. Detailed discussion on 
 the double pole structure of the $\Lambda(1405)$ in chiral unitary 
 approaches can be found in Refs.~\cite{jido2,Jido:2010ag}.} 
of the $\Lambda(1405)$ couples strongly to
the $\pi\Sigma$ channel~\cite{jido2,jido3}. Thus, this article is devoted to further clarification of the properties of this new $N^{*}$ state. To do that we follow the coupled-channel Faddeev approach 
developed in Ref.~\cite{mko2} but taking into account coupled channels of $K\pi\Sigma$ 
and $K\pi\Lambda$ together with $K\bar{K}N$. 

Let us briefly explain our formulation to study three-body coupled channels 
of two mesons and one baryon with $J^{P}=1/2^{+}$. 
Paying special attention to the dynamical generation of the $\Lambda(1405)$
in the $\bar KN$ and $\pi\Sigma$ subsystems, we consider the $K\bar K N$, 
$K\pi\Sigma$ and $K\pi\Lambda$ channels with total charge zero, namely, in the charge base,
$K^{0}K^{-}p$, $K^{0}\bar{K}^{0}n$, $K^{0}\pi^{0}\Sigma^{0}$, 
$K^{0}\pi^{+}\Sigma^{-}$, $K^{0}\pi^{-}\Sigma^{+}$, $K^{0}\pi^{0}\Lambda$, 
$K^{+}K^{-}n$, $K^{+}\pi^{0}\Sigma^{-}$, $K^{+}\pi^{-}\Sigma^{0}$, 
$K^{+}\pi^{-}\Lambda$, and calculate the three-body $T$ matrix for the different transitions. 

To determine the three-body $T$ matrix we follow the formalism developed in 
Refs.~\cite{mko1,mko2,mko3,mko7} which is based on the Faddeev equations \cite{Fa}. 
In terms of the Faddeev partitions, $T^1$, $T^2$ and $T^3$, the three-body 
$T$-matrix is written as 
\begin{equation}
T=T^1+T^2+T^3 . \label{T}
\end{equation}
In our formalism these partitions are expressed as \cite{mko1,mko2,mko7}
\begin{equation}
T^i =t^i\delta^3(\vec{k}^{\,\prime}_i-\vec{k}_i) + \sum_{j\neq i=1}^3T_R^{ij}, 
\label{Ti}
\end{equation}
for $i=1,2,3$
with $\vec{k}_{i}$ ($\vec{k}^\prime_{i}$) being the initial (final) momentum of the particle $i$ and $t^{i}$, $i=1,2,3$, the two-body $t$-matrix which describes the interaction for the $(jk)$ pair of the system, $j \neq k\neq i=1,2,3$. In our approach, this two-body $t$-matrix is calculated by solving the Bethe-Salpeter equation with the potential obtained from chiral Lagrangians~\cite{oller1,oller2, eo1,oller4,jido2}.
Namely we consider all possible two-body channels of meson ($\pi$, $\eta$, $K$, $\bar K$) and baryon ($N$, $\Lambda$, $\Sigma$, $\Xi$) which couple to $\bar KK$, $\pi K$, $\bar KN$, $K\Sigma$ and $KN$, but except for the $\eta\eta$ channel, which is not important in $\pi\pi$ and $\bar KK$ dynamics~\cite{oller1}.

In Eq.~(\ref{Ti}), the $T^{ij}_{R}$ partitions include all the different contributions to the three-body $T$  matrix in which the last two interactions are given in terms of the two-body $t$-matrices $t^j$ and $t^i$, respectively, and satisfy the following set of coupled equations
\begin{equation}
T^{\,ij}_R = t^ig^{ij}t^j+t^i\Big[G^{\,iji\,}T^{\,ji}_R+G^{\,ijk\,}T^{\,jk}_R\Big], 
  \label{Trest}
\end{equation}
for  $i\ne j, j\ne k = 1,2,3$.

In Eq.~(\ref{Trest}), $g^{ij}$'s correspond to the three-body Green's function of the system and its elements are defined as
\begin{eqnarray}
\lefteqn{
g^{ij} (\vec{k}^\prime_i, \vec{k}_j)=\Bigg(\frac{N_{k}}{2E_k(\vec{k}^\prime_i+\vec{k}_j)}\Bigg)}&& \nonumber \\
&& \times \frac{1}{\sqrt{s}-E_i
(\vec{k}^\prime_i)-E_j(\vec{k}_j)-E_k(\vec{k}^\prime_i+\vec{k}_j)+i\epsilon},
\end{eqnarray}
with $N_{k}=1$ for mesons and $N_{k}=2M_{k}$ for baryons with baryon mass $ M_{k}$, and $E_{l}$, $l=1,2,3$, is the energy of the particle $l$. The $G^{ijk}$ matrix in Eq. (\ref{Trest}) represents  a loop function of three-particles and it is written as
\begin{equation}
G^{i\,j\,k}  =\int\frac{d^3 k^{\prime\prime}}{(2\pi)^3}\tilde{g}^{ij} \cdot F^{i\,j \,k}
\end{equation}
with the elements of  $\tilde{g}^{ij}$ being 
\begin{eqnarray}
\lefteqn{
\tilde{g}^{ij} (\vec{k}^{\prime \prime}, s_{lm}) = \frac{N_l}
{2E_l(\vec{k}^{\prime\prime})} \frac{N_m}{2E_m(\vec{k}^{\prime\prime})} } && \nonumber \\
&& \quad \times
\frac{1}{\sqrt{s_{lm}}-E_l(\vec{k}^{\prime\prime})-E_m(\vec{k}^{\prime\prime})
+i\epsilon},
\label{eq:G} 
\end{eqnarray}
for $i \ne l \ne m$, 
and the matrix $F^{i\,j\,k}$, with explicit variable dependence, is given by 
\begin{eqnarray}
\lefteqn{
F^{i\,j\,k} (\vec{k}^{\prime \prime},\vec{k}^\prime_j, \vec{k}_k,  s^{k^{\prime\prime}}_{ru})=  } && \nonumber \\
&& t^{j}(s^{k^{\prime\prime}}_{ru}) g^{jk}(\vec{k}^{\prime\prime}, \vec{k}_k)
\Big[g^{jk}(\vec{k}^\prime_j, \vec{k}_k) \Big]^{-1}
\Big[ t^{j} (s_{ru}) \Big]^{-1},  \label{offac}
\end{eqnarray}
for $ j\ne r\ne u=1,2,3$.
In Eq. (\ref{eq:G}), $\sqrt{s_{lm}}$ is the invariant mass of the $(lm)$ pair and can be calculated in terms of the external variables. The upper index $k^{\prime\prime}$ in the invariant mass $s^{k^{\prime\prime}}_{ru}$ of Eq.~(\ref{offac}) indicates its dependence on the loop variable, as it was shown in Ref. \cite{mko2}. 

The main advantage of Eq.~(\ref{Trest}) is that they are algebraic coupled equations and not integral equations as it was shown in Refs. \cite{mko1,mko2}. In these works, for the first time, cancellation between the contribution of the off-shell parts of the chiral two-body $t$-matrices to the three-body diagrams and the contact term with three particles in the initial and final state, whose origin is in the chiral Lagrangian used to describe the interaction, was found analytically (see Refs. \cite{mko1,mko2,mko3} for more details).

In terms of the $T_{R}^{ij}$ partitions (Eq. (\ref{Trest})), 
the expression for the full three-body $T$-matrix can be obtained combining Eq.~\eqref{T} and Eq.~\eqref{Ti}.
Nontrivial three-body dynamics appears in 
\begin{eqnarray}
T_{R} \equiv \sum_{i=1}^3\sum_{j\neq i=1}^{3}T^{ij}_{R} . \label{ourfullt}
\end{eqnarray}

This amplitude is a function of the total three-body energy, $\sqrt s$, and the 
invariant mass of  the particles 2 and 3, $\sqrt{s_{23}}$. The other invariant masses, $\sqrt{s_{12}}$ and $\sqrt{s_{31}}$ can be obtained in terms of  $\sqrt s$ and $\sqrt{s_{23}}$, as it was shown in Ref. \cite{mko2,mko3}.
To present our results, we have chosen $\sqrt{s}$ and the invariant mass of one of the two-body subsystems, $\sqrt{s_{ij}}$ .
All the matrices in  Eq.~(\ref{Trest}) are projected 
in S-wave, thus the quantum numbers of the three-body system and, hence, the resulting resonances are $J^{\pi}=1/2^{+}$.

Let us discuss the results obtained for the three-body amplitude $T_{R}$. 
Our discussion concentrates on showing $|T_{R}|^2$ for real values of 
$\sqrt s$ and of the invariant mass of one of the two-body subsystems, in concrete, $\sqrt{s_{\bar{K}N}}$ and $\sqrt{s_{K\bar{K}}}$. 
We have solved Eq.~(\ref{Trest}) in the charge base, thus,  to study
the existence of a three-body $N^*$ resonance around $1920$ MeV we have to project $T_{R}$ on the isospin base with total isospin $I=1/2$.
Our interest is to examine the possibility of existence of a $N^{*}$ resonance which appears as a  
$K\bar{K}N$ bound state when the $\bar{K}N$ subsystem and its coupled 
channels generate the $\Lambda(1405)$, as pointed out in Ref.~\cite{jido}.
For this purpose, we specify also the isospin of the two-body subsystem.
First, we calculate the three-body $T_{R}$ matrix for the $K\bar{K}N$ channel projected 
on  total isospin $I=1/2$ and the $\bar KN$ subsystem with $I_{\bar KN}=0$,
which we represent as $(T^{K\bar{K}N}_{R})^{(I=1/2,I_{\bar KN}=0)}=\langle I=1/2, I_{\bar{K}N}=0| T^{K\bar{K}N}_{R}|I=1/2, I_{\bar{K}N}=0\rangle$.
In addition, to understand the characteristics of this $N^{*}$ further, we also show 
the $T_{R}$ matrix projected on $I=1/2$ with the $\bar KK$ subsystem in isospin one,  $I_{\bar KK}=1$, which is denoted by
$(T^{K\bar{K}N}_{R})^{(I=1/2,I_{\bar KK}=1)}=\langle I=1/2, I_{\bar{K}K}=1| T^{K\bar{K}N}_{R}|I=1/2, I_{\bar{K}K}=1\rangle$.
In the $\bar KK$ two-body subsystem with $I_{\bar KK}=1$, $a_{0}(980)$ is 
dynamically generated (Ref. \cite{oller1,oller2}).

\begin{figure}[t!]
\centering
\begin{tabular}{cc}
\hspace{-1cm}
\vspace{-0.38cm}
\resizebox{0.3\textwidth}{!}{\includegraphics[width=\textwidth]{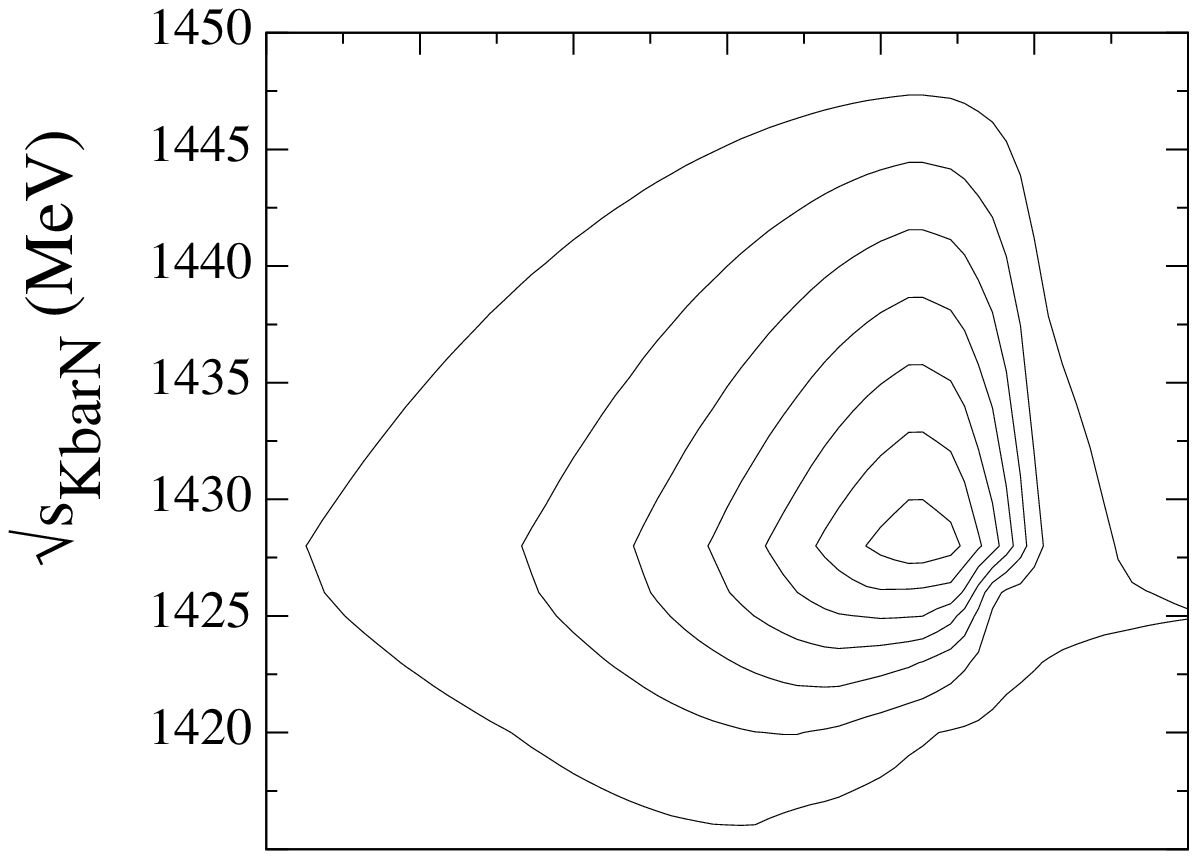}}\\
\hspace{-1cm}
\resizebox{0.3\textwidth}{!}{\includegraphics[width=\textwidth]{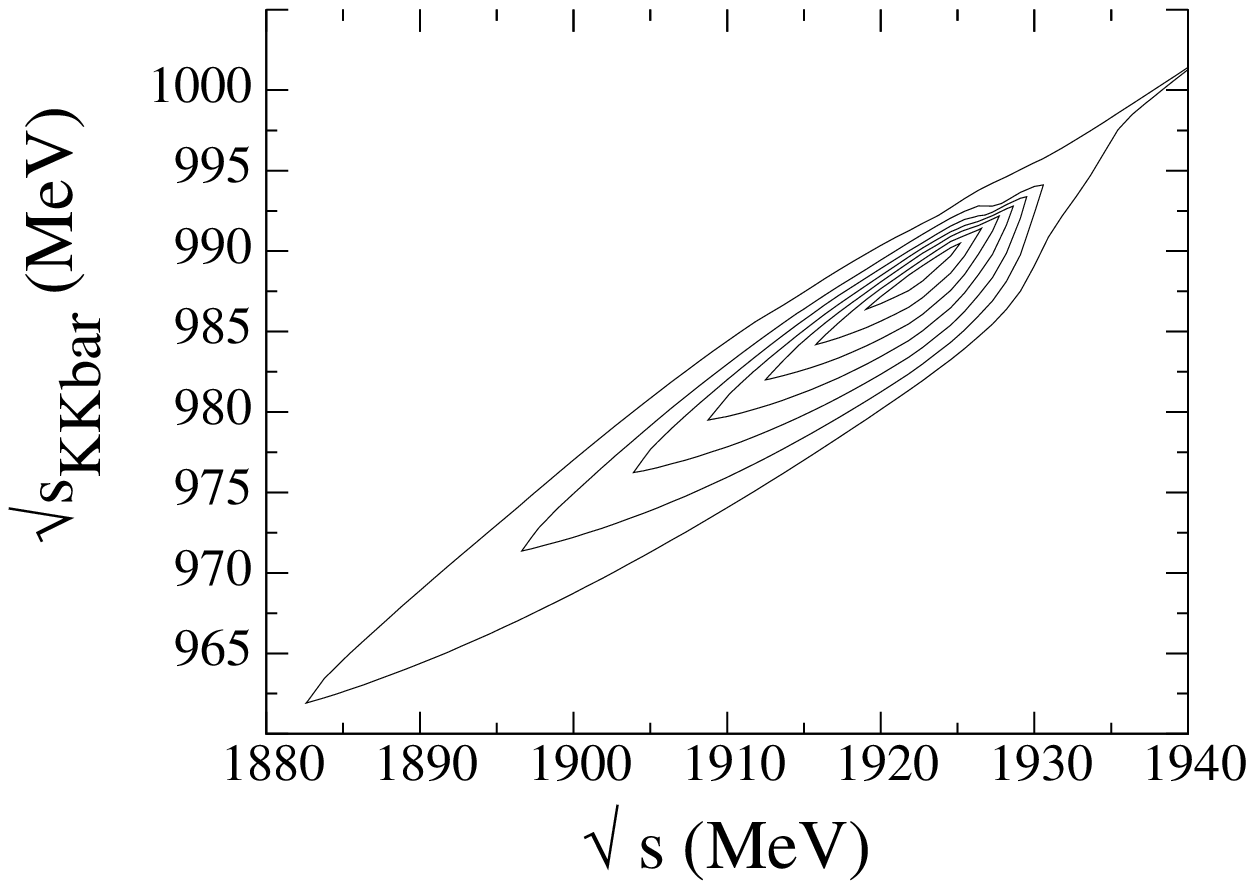}}
\end{tabular}

\caption{Contour plots of the three-body squared amplitude $|T_{R}|^2$ 
for the $N^{*} $ resonance in the $K\bar{K}N$ system
as functions of the total three-body energy, $\sqrt s$, and
the invariant mass of the $\bar KN$ subsystem with $I_{\bar KN}=0$
(upper panel) or the invariant mass of the $\bar KK$ subsystem with 
$I_{\bar KK}=1$ (lower panel).} \label{N1}
\end{figure}

In Fig.~\ref{N1} we show the contour plots corresponding to the three-dimensional plots of the squared three-body $T_{R}$ matrix, $(T^{K\bar{K}N}_{R})^{(I=1/2,I_{\bar KN}=0)}$ (upper panel) and $(T^{K\bar{K}N}_{R})^{(I=1/2,I_{\bar KK}=1)}$ (lower panel)
plotted as functions of the total energy of the three-body system, $\sqrt{s}$, and the $\bar{K} N$ invariant mass, $\sqrt{s_{\bar{K}N}}$, and the $\bar{K} K$ invariant mass, $\sqrt{s_{\bar{K}K}}$, respectively.  As it can be seen in the upper panel, a peak in the squared amplitude is obtained around $\sqrt{s}\sim1922$ MeV when the $\bar{K}N$ subsystem in isospin zero has an invariant mass close to 1428 MeV. In the lower panel, 
the peak shows up when the invariant mass of the $K\bar{K}$ subsystem is around 987 MeV.

We also find the $N^{*}$ resonance at the same value of $\sqrt s$ in the $T_{R}$ matrices 
for different isospin combinations of the $\bar KN$ and $\bar KK$ subsystems. 
For the case in which the $\bar KN$ subsystem is in isospin 1,  $I_{\bar KN}=1$, the
$(T^{K\bar{K}N}_{R})^{(I=1/2,I_{\bar KN}=1)}$ matrix shows a less pronounced peak structure 
for the $N^{*}$, due to the fact that the projected $T_{R}$ matrix on $I_{\bar KN}=1$ has 
tiny contributions of the $\Lambda(1405)$ in the intermediate states. The ratio of the
$|T_{R}|^2$ matrices with $I_{\bar KN}=1$ and $I_{\bar KN}=0$ at the resonance point,
$(\sqrt{s},\sqrt{s_{\bar KN}})=(1923,1428)$ MeV,
is found to be a tiny value, $\sim$ 0.008. Also, for the $\bar KK$ subsystem, the ratio of $|T_{R}|^2$
with $I_{\bar KK}=0$ and $I_{\bar KK}=1$ at $(\sqrt{s},\sqrt{s_{\bar KK}})=(1923,987)$ MeV is $\sim$ 1. Although the magnitude for these two matrix elements with $I_{\bar KK}=0$ and $I_{\bar KK}=1$ is  very similar, it does not mean that the fraction of the $I_{\bar KK}=0$ and 
$I_{\bar KK}=1$ components in the $N^*$ state is similar, since this fraction depend 
on the isospin configuration of the $\bar KN$ subsystem. 
Group theory tells us that, in  case the $\bar KN$ subsystem has purely 
$I_{\bar KN}=0$, the ratio of the $I_{\bar KK}=1$ and $I_{\bar KK}=0$
components is 3 to 1 for total $I=1/2$. Since, in the present case,  the $\bar K N$ pair dominantly has 
$I_{\bar K N}=0$, the $I_{\bar KK}=1$ component is favored in the  $N^{*}$ state. This implies that the $N^{*}$ resonance contains mostly $a_{0}(980)$
in the $\bar KK$ subsystem with $I_{\bar KK}=1$ and less contribution from 
$f_{0}(980)$ with $I_{\bar KK}=0$.  The attraction in the $K\bar{K}$ and $\bar{K}N$ subsystems is strong enough to compensate the repulsion in the $K N$
subsystem and form a bound state.

It is well known that the $\bar{K}N$ interaction and coupled channels generate the $\Lambda(1405)$ state, with a double pole structure \cite{jido2}: there is a pole around 1390 MeV, which couples strongly to $\pi\Sigma$, and another one around 1426 MeV, which couples dominantly to $\bar{K}N$.  Similarly, the $K\bar{K}$ interaction and coupled channels generate the $\sigma(600)$, $f_0(980)$, and $a_{0}(980)$ resonances \cite{oller1,oller2}. Therefore, the fact that the peak of the three-body $T_{R}$ matrix around 1920 MeV gets generated when the $\bar{K}N$ subsystem has an invariant mass close to 1428 MeV with $I_{\bar{K}N}=0$ and, at the same time, the $K\bar{K}$ subsystem has an invariant mass of around 987 MeV with $I_{K\bar{K}}=1$ is indicating that the $N^{*}$ resonance obtained has an important $K\Lambda(1405)$ and $a_{0}(980)N$ components. 
This result is consistent with what was found in the variational method~\cite{jido}.

\begin{figure}[t!]
\centering
\begin{tabular}{cc}
\resizebox{0.23\textwidth}{!}{\includegraphics[width=\textwidth]{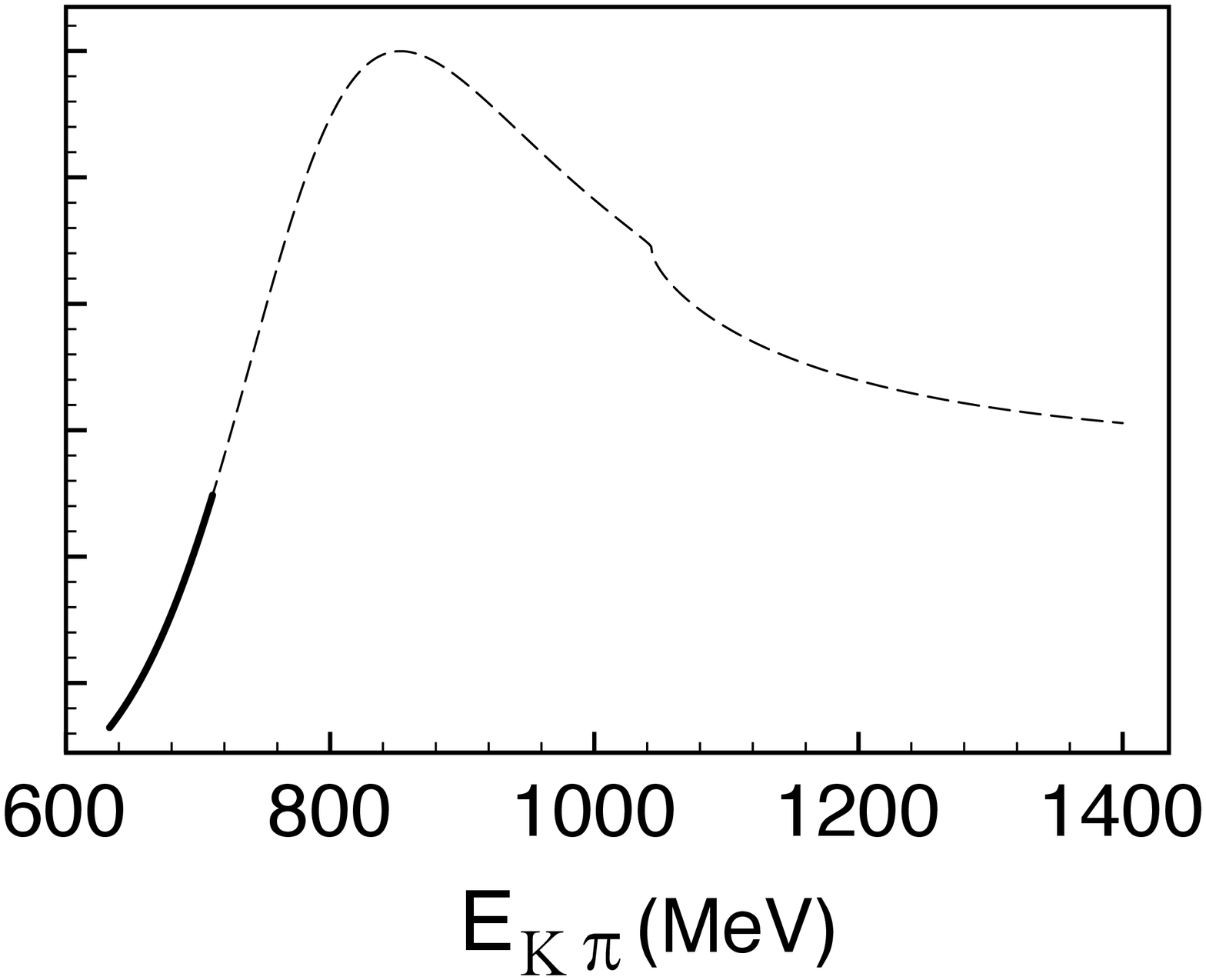}}
\resizebox{0.23\textwidth}{!}{\includegraphics[width=\textwidth]{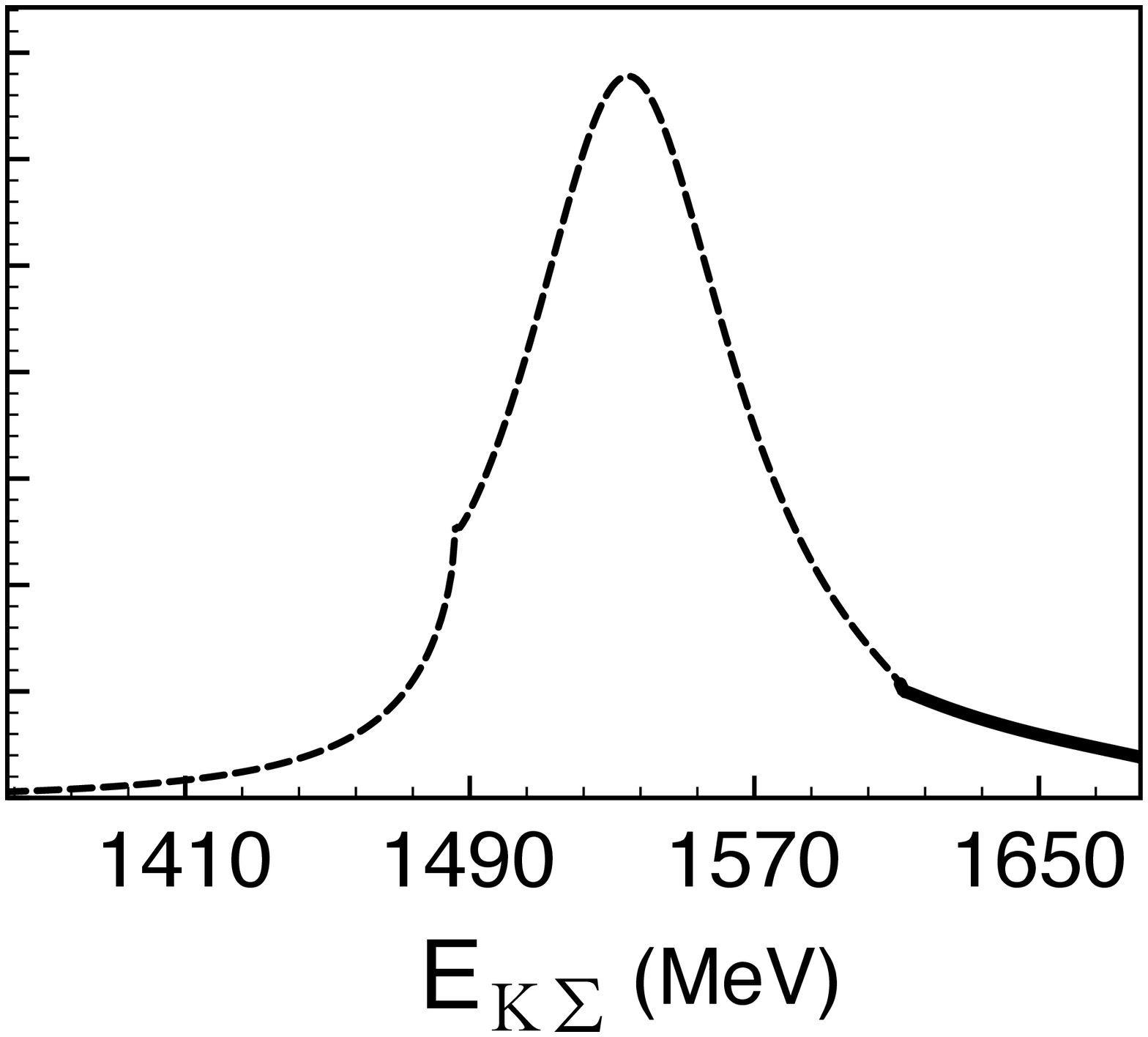}}
\end{tabular}
\caption{Dashed Lines: (Left)  $K\pi\to K\pi$ $t$ matrix with isospin $1/2$. The $\kappa$ resonance is generated around 850 MeV. (Right) $t$ matrix for the $K\Sigma\to K\Sigma$ transition with isospin $1/2$, which shows the presence of the $N^{*}(1535)$. Solid line:  Energy range used in the three-body calculation for the $K\pi$ interaction (Left) and for the $K\Sigma$ interaction (Right). The units are arbitrary.}\label{Kpi}
\end{figure}



We also find that the contribution of the $K\pi\Sigma$ and $K\pi\Lambda$ channels to the three-body $T_{R}$ matrix does not alter the peak position and width found in the case in which only the $K\bar{K}N$ channel was considered. Therefore, solving the equations only with the $K\bar{K}N$ channel alone gives the same results as the ones obtained in the coupled channel approach. The result that the $K\pi\Sigma$ channel does not seem to play an important role in the dynamical generation of the $K\bar{K}N$ bound state can be understood by considering the fact that while in the $\pi\Sigma$ system we have the presence of the $\Lambda(1405)$ resonance, the energy range relevant for the $K\pi$ and $K\Sigma$ interactions in the three-body calculation (solid line in Fig. \ref{Kpi}, respectively) is far from the energy in which the $\kappa(850)$ and $N^{*}(1535)$ get dynamically generated, which are the only resonances generated with the unitary chiral approach in the respective systems~\cite{oller1,oller2,inoue}, and, thus, these interactions are weaker as compared to those of $K\bar{K}$ and $\bar{K}N$ in the $K\bar{K}N$ channel. We have also solved Eq. (\ref{Trest}) without the inclusion of the $K\bar{K}N$ channel and we have  found a signal for the $N^*(1910)$ in the $K\pi \Sigma$ channel\footnote{The energy position as well as the $\pi\Sigma$ invariant mass for which the peak gets generated is a bit different from the case where the $K\bar{K}N$ channel is considered: $\sqrt{s}\sim$ 1915 MeV, $\sqrt{s_{\pi\Sigma}}\sim1415$ MeV. This fact could be related to the double pole structure of the $\Lambda(1405)$. }, although it is much weaker than the one obtained when the $K\bar{K}N$ channel was added. Similar situation occurs for the $K\pi\Lambda$ channel. Therefore, the $N^*(1910)$ is made mainly by $K\bar{K}N$, where one has the presence of the $\Lambda(1405)$ in the $\bar{K}N$ interaction as well as the $a_{0}(980)$ in the $K\bar{K}$ interaction. 

Recently, it has been pointed out in Ref.~\cite{Ikeda:2010tk} that 
two resonance poles are found in the $\bar KNN$ system with total isospin $I=1/2$ and 
$J^{P}=1/2^{+}$ between the $\bar KNN$ and $\pi \Sigma N$ threshold, when
a Weinberg-Tomozawa energy dependent two-body interaction is used for a
three-body Faddeev calculation: one pole is located moderately below 
the $\bar KNN$ threshold with a narrow width, while the other one appears above 
the $\pi \Sigma N$ threshold with a substantially large width.
These poles are associated with the two-pole nature 
of the $\Lambda(1405)$ generated in the $\bar KN$ and $\pi \Sigma$ subsystems. 
We have also looked for another resonance state associated with the lower 
$\Lambda(1405)$ pole, but we could not find any signal for such a deep state in the $T_{R}$
matrix evaluated with real values of $\sqrt s$. This means either that 
there is no such a resonance state or that 
there is a resonance state having such a  large width that 
the resonance contribution cannot be seen in the real axis of $\sqrt s$.

To summarize, we have studied the $K\bar{K}N$, $K\pi\Sigma$ and $K\pi\Lambda$ systems by solving the Faddeev equations in a coupled channel approach. The input two-body $t$-matrices have been obtained by using potentials from  chiral Lagrangians and solving the Bethe-Salpeter equations in a unitary coupled channel approach.  We have found the contribution of the $K\pi\Sigma$ and $K\pi\Lambda$ channels to the three-body $T$ matrix to be negligible around a total energy for the three-body system close to $1920$ MeV and, thus, one could solve the Faddeev equations considering only the $K\bar{K}N$ channel. The resolution of that equations has lead to the dynamical generation of a $N^{*}$ resonance around 1920 MeV with $J^{\pi}=1/2^{+}$, as  was predicted in \cite{jido} and found in \cite{mko1}. The resonance is generated when the $\bar{K}N$ subsystem is resonating as the $\Lambda(1405)$ and, at the same time, the $K\bar{K}$ interaction generates the $a_{0}(980)$ resonance.  

The authors  thank Dr.~Hyodo for his useful comments, a part of
which motivated this work, and Professors~E.~Oset, K.~Kanada-Eny'o
and Dr.~K.~P.~Khemchandani for collaboration in the previous works
on which the present one is based.
The work of A.~M.~T.~is supported by  
the Grant-in-Aid for the Global COE Program ``The Next Generation of Physics, 
Spun from Universality and Emergence" from the Ministry of Education, Culture, 
Sports, Science and Technology (MEXT) of Japan.
This work is supported in part by
the Grant for Scientific Research (No.~22105507) from 
MEXT of Japan.
A part of this work was done in the Yukawa International Project for 
Quark-Hadron Sciences (YIPQS).

\end{document}